\documentclass[preprint]{aastex631}

\usepackage{amsmath}
\usepackage{bm}
\usepackage{verbatim}

\shorttitle{High-velocity stars}
\shortauthors{Liao et al.}

\begin{document}

\title{The origin of High-velocity stars considering the impact of the Large Magellanic Cloud}

\correspondingauthor{Cuihua Du}
\email{ducuihua@ucas.ac.cn}

\author{Jiwei Liao}
\affiliation{School of Astronomy and Space Sciences, University of Chinese Academy of Sciences, Beijing 100049, P.R. China}

\author{Cuihua Du}
\affiliation{School of Astronomy and Space Sciences, University of Chinese Academy of Sciences, Beijing 100049, P.R. China}

\author{Mingji Deng}
\affiliation{School of Astronomy and Space Sciences, University of Chinese Academy of Sciences, Beijing 100049, P.R. China}

\author{Dashuang Ye}
\affiliation{School of Astronomy and Space Sciences, University of Chinese Academy of Sciences, Beijing 100049, P.R. China}

\author{Hefan Li}
\affiliation{Key Laboratory of Optical Astronomy, National Astronomical Observatories, Chinese Academy of Sciences, Beijing 100012, P.R.China}

\author{Yang Huang}
\affiliation{School of Astronomy and Space Sciences, University of Chinese Academy of Sciences, Beijing 100049, P.R. China}
\affiliation{Key Laboratory of Optical Astronomy, National Astronomical Observatories, Chinese Academy of Sciences, Beijing 100012, P.R.China}

\author{Jianrong Shi}
\affiliation{Key Laboratory of Optical Astronomy, National Astronomical Observatories, Chinese Academy of Sciences, Beijing 100012, P.R.China}
\affiliation{School of Astronomy and Space Sciences, University of Chinese Academy of Sciences, Beijing 100049, P.R. China}

\author{Jun Ma}
\affiliation{Key Laboratory of Optical Astronomy, National Astronomical Observatories, Chinese Academy of Sciences, Beijing 100012, P.R.China}
\affiliation{School of Astronomy and Space Sciences, University of Chinese Academy of Sciences, Beijing 100049, P.R. China}

\begin{abstract}

Utilizing astrometric parameters sourced from \textit{Gaia} Data Release 3 and radial velocities obtained from various spectroscopic surveys, we identify 519 high-velocity stars (HiVels) with a total velocity in the Galactocentric restframe greater than 70\% of their local escape velocity under the {\tt\string Gala} {\tt\string MilkyWayPotential}. Our analysis reveals that the majority of these HiVels are metal-poor late-type giants, and we show 9 HiVels that are unbound candidates to the Galaxy with escape probabilities of 50\%. To investigate the origins of these HiVels, we classify them into four categories and consider the impact of the Large Magellanic Cloud (LMC) potential on their backward-integration trajectories. Specifically, we find that one of the HiVels can track back to the Galactic Center, and three HiVels may originate from the Sagittarius dwarf spheroidal galaxy (Sgr dSph). Furthermore, some HiVels appear to be ejected from the Galactic disk, while others formed within the Milky Way or have an extragalactic origin. Given that the LMC has a significant impact on the orbits of Sgr dSph, we examine the reported HiVels that originate from the Sgr dSph, with a few of them passing within the half-light radius of the Sgr dSph.

\end{abstract}


\section{Introduction} \label{1}

The existence of high-velocity stars (HiVels) implies the presence of extreme dynamics. In the literature, these stars are typically divided into four categories: hypervelocity stars (HVSs), hyper-runaway stars (HRSs), runaway stars (RSs), and high velocity halo stars  \citep{Li.Yinbi2021,Quispe-Huaynasi2022}. The first two types of HiVels are unbound stars, with speeds exceeding the escape velocity of the Milky Way (MW). The classic origin of HVSs, as theorized by \citet{Hill1988}, involves a close encounter between a stellar binary and a massive black hole (MBH). A star that is ejected by a MBH and has an unbound velocity is defined as an HVS by \citet{Brown2015}. The Southern Stellar Stream Spectroscopic Survey (S$^{5}$) discovered the fastest HVS \citep{Koposov2020}, which is an A-type star with a total velocity of 1755\,km\,s$^{-1}$. This star was ejected by the MBH at the Galactic Center, providing strong evidence for the Hills mechanism. HRSs are another type of unbound HiVels and are also considered as RSs with extreme velocities. One of the most plausible mechanisms for generating HRSs is the thermonuclear explosion in double white dwarf binaries \citep{Shen2018}. The first confirmed HRS is HD 271791 originating from the outer disk \citep{Heber2008}, with a total velocity of 630\,km\,s$^{-1}$. 

\par The last two types of HiVels are stars that are bound by the Galactic potential. RSs were initially introduced as O- and B-type stars with high velocities \citep{Blaauw1961}. It is believed that RSs form in the disk and then are ejected into the halo. There are generally two main mechanisms for the formation of RSs: supernova explosions in stellar binary systems \citep[e.g.,][]{Blaauw1961,Portegies2000,Wang2009,Wang2013} and dynamical encounters resulting from multi-body interactions in dense stellar systems \citep[e.g.,][]{Bromley2009,Gvaramadze2009}. High-velocity halo stars are another type of bound HiVels that could be an in-situ population of stars formed within the MW \citep{DiMatteo2019,Belokurov2020} or an extragalactic source.  Numerical simulations by \citet{Abadi2009} have shown that disrupted dwarf galaxies may contribute halo stars with high velocities. In addition to the mechanisms mentioned above, the dense environments in the cores of globular clusters facilitate many strong dynamical encounters among stellar objects \citep{Cabrera2023}. HiVels could also be ejected through the close interaction of a globular cluster and a supermassive black hole \citep{Capuzzo-Dolcetta2015}, or through the formation of three-body binaries in globular clusters \citep{Weatherford2023}.

\par Recently, a significant number of candidate HiVels originating from the Sagittarius dwarf spheroidal galaxy (Sgr dSph) have been identified. The first candidate HVS (J1443+1453) that likely originated from the Sgr dSph was discovered by \citet{Huang2021}. Subsequently, \citet{Lihf2022} reported 60 candidate HiVels (including 2 HVS candidates) that may also come from the Sgr dSph, and \citet{LiQZ2023} found 15 extreme velocity stars that have had close encounters with the Sgr dSph. However, both \citet{Huang2021} and \citet{LiQZ2023} did not consider the impact of the Large Magellanic Cloud (LMC), and \citet{Lihf2022} disregarded the dynamical friction from the Milky Way (MW) on the LMC. It is important to mention that the HVS3 (or HE 0437-5439) has a 40\% probability of passing within 5\,kpc of the LMC, taking into account the LMC's influence \citep{Edelmann2005,Erkal2019}. It is fortunate that the \textit{Gaia} Data Release 3 (\textit{Gaia} DR3) provides precise proper motion measurements for HVS3. Consequently, it is necessary to investigate the origin of these HiVels. 

\par The aim of the work is to search for more HiVels, and determine their origins considering the impact of the LMC. This paper is organized as follows: Section \ref{2} provides a description of the observation sample. In section \ref{3}, we obtain 519 HiVels with a total velocity in the Galactocentric restframe greater than 70\% of the local escape velocity. Section \ref{4} focuses on the analysis of the origin of these HiVels. Finally, a summary is provided in Section \ref{5}.

\section{Observation sample}
\label{2} 

\par
\begin{figure}[ht!]
\plotone{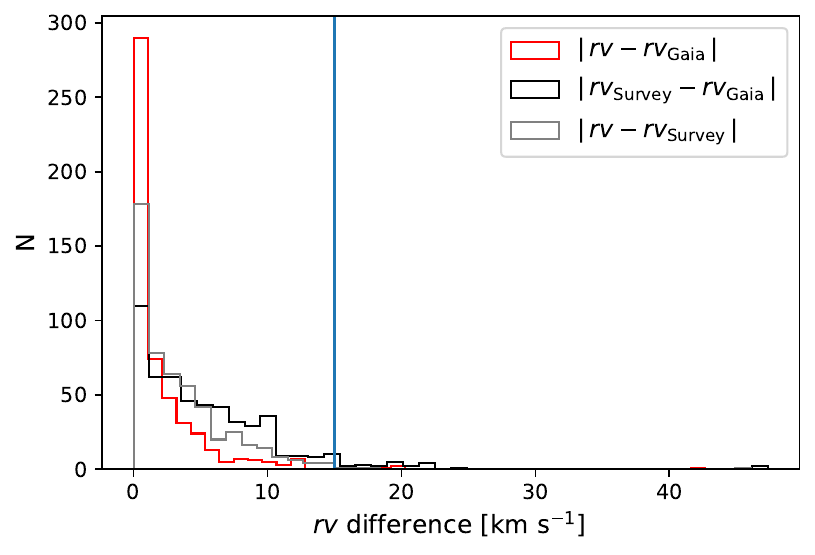}
\caption{
Distribution of radial velocity differences ($\lvert rv-rv_\mathrm{Gaia} \rvert$, $\lvert rv_\mathrm{Survey}-rv_\mathrm{Gaia} \rvert$, $\lvert rv-rv_\mathrm{Survey} \rvert$) for the \textbf{519} HiVels. The corrected radial velocities, the radial velocities from Gaia, and the radial velocities from other surveys are represented by $rv$, $rv_\mathrm{Gaia}$, and $rv_\mathrm{Survey}$, respectively. For reference, a vertical blue line is plotted at a difference of 15 \,km\,s$^{-1}$. \label{fig4} } 
\end{figure}

  The \textit{Gaia} DR3 provides astrophysical parameters for 470 million stars \citep{GaiaDR3}, 34 million radial velocities (RVs) \citep{Katz2023}. To select stars with accurately measured parallax, proper motion and spectroscopic data, we applied the following criteria: {\tt\string RUWE}\,$<1.4$, $\rm \varpi-\varpi_\mathrm{ZP}$\,$>$\,$0$, $\varpi-\varpi_\mathrm{ZP}$\,\textgreater\,5$\sigma_\varpi$ and S/N\,$>$\,$10$.  {\tt\string RUWE} is Renormalized Unit Weight Error to assess the goodness-of-fit of the astrometric solution for a particular star\citep{Lindegren2021a}. We use the approach described in \citet{Lindegren2021b} to correct \textit{Gaia} DR3 parallaxes by the parallax zero point $\varpi_\mathrm{ZP}$.  
To obtain precise RVs, we cross-matched the \textit{Gaia} DR3 catalog with other large-scale Galactic Surveys, such as GALAH DR3 \citep{GALAH2015,GALAH2021}, LAMOST DR10 \citep{LAMOSTcui2012,LAMOSTzhao2012}, RAVE DR6 \citep{RAVE2020}, APOGEE DR17 \citep{APOGEE2017}, and S$^{5}$ \citep{S52019}). 
The weighted mean RV was computed using the following equation:
\begin{equation}
rv=\frac{\sum_{i}^{N}w_{i}rv_{i}}{\sum_{i}^{N}w_{i}},
\end{equation}
where $rv$ represents the calculated radial velocity, \textbf{$w_{i}=1/\sigma_{i}^{2}$} is the weight corresponding to each measurement with uncertainty $\sigma_{i}$, and $N$ is the total number of measurements. We removed stars for which $\lvert rv-rv_\mathrm{Gaia} \rvert\geqslant3\sigma_\mathrm{Gaia}$. In the context of this analysis, $rv$, $rv_\mathrm{Gaia}$, and $rv_\mathrm{Survey}$ correspond to the corrected radial velocity, the radial velocity from Gaia, and the radial velocity from other surveys, respectively. Figure \ref{fig4} demonstrates that the difference in radial velocities among the \textbf{519} HiVels is minimal, with a majority falling within 15\,km\,s$^{-1}$. Using the \textit{Gaia}'s parameters such as {\tt\string phot\_variable\_flag}, and {\tt\string classprob\_dsc\_combmod\_star} flag which represents the probability of being a single star (but not a white dwarf) from DSC-Combmod (which classifies objects using BP/RP spectrum, photometry and astrometry features), we removed all variable stars and the 519 HiVels are single stars.

\par  Adopting the same method as \citet{Du2019}, we use Bayesian analysis to determine heliocentric distances ($d$) and velocities in the R.A. and Decl. direction ($\nu_{\alpha}$ and $\nu_{\delta}$). The posterior consists of likelihood with a three-dimensional Gaussian distribution and prior with the three-parameter generalized Gamma distribution \citep{Bailer-Jones2021}:
\begin{equation}
P(\theta |\chi )\propto e^{-\frac{1}{2}(\chi -m(\theta ))^TC_{\chi}^{-1}(\chi -m(\theta ))}\frac{1}{\Gamma (\frac{\beta +1}{\alpha })}\frac{\alpha}{L^{\beta +1}}r^{\beta}e^{-(\frac{r}{L})^\alpha},
\end{equation}
where $\theta$\,=\,$\left[ d,\nu_{\alpha },\nu_{\delta}\right ]$, $\chi$\,=\,$\left[ \varpi,\mu _{\alpha*},\mu _{\delta }\right ]$, $m$\,=\,$[ 1/d+\varpi_\mathrm{zp},\nu_{\alpha }/kd,\nu_{\delta}/$
$kd ]$, $k$\,=\,4.74047, and $C_{\chi}$ is covariance matrix.

\par We implement the Markov Chain Monte Carlo (MCMC) sampler emcee \citep{emcee} to draw samples from the posterior probability, and generate 4000 Monte Carlo (MC) realizations for each object (stars, dwarf galaxies, and globular clusters). Following the approach outlined in \citet{Liao2023}, we adopt the position of the Sun to the Galactic Centre $( x_{\odot}, y_{\odot},z_{\odot})=(8.122,0,0.0208)$\,kpc \citep{GRAVITY2018A&A...615L..15G,Bennett2019} and solar velocity of $( U_{\odot}, V_{\odot},W_{\odot})= (11.1,245,7.25)$\,km\,s$^{-1}$ \citep{Sch2010,McMillan2017} to derive 3D velocities $(U, V, W)$ \citep{Johnson1987} and position of each star in the Cartesian Galactocentric coordinate system \citep{Juri2008}.

\section{HiVels sample}\label{3}

\par In order to determine the origin of the HiVels, we rewind them in the combined presence of the gravitational field of the MW and LMC. For the MW, we adopt the {\tt\string Gala} potential {\tt\string MilkyWayPotential} \citep{gala}, including the Hernquist \citep{Hernquist1990} bulge and nucleus, the Miyamoto-Nagai \citep{Miyamoto1975} stellar disk, and the Navarro-Frenk-White \citep{NFW1996} dark matter halo. For the LMC, we consider a Plummer profile with a mass of $1.38 \times 10^{11} M_\odot$ \citep{Erkal2019mass} and dynamical friction from the MW on the LMC taking the prescription from \citet{Petts2016}. We have set the half-mass radius to 5 kpc for definiteness. The scale radius is set at 17\,kpc, computed based on an enclosed mass of $1.7 \times 10^{10} M_\odot$ within 8.7\,kpc \citep{van2014}. The analysis is conducted with the LMC's position, radial velocity and proper motions from \citet{Erkal2019}. The dynamical friction and LMC's gravitational potential use the {\tt\string ChandrasekharDynamicalFrictionForce} function and {\tt\string MovingObjectPotential} function implemented in {\tt\string galpy} \citep{bovy2015}, respectively. We then integrate the backward orbit \textbf{using {\tt\string galpy}} over a total time of 3 Gyr with a time step of 0.1 Myr . 

\par If a star is ejected from the Galactic disk, its backward-integral trajectory passes through the Galactic midplane ($Z_\mathrm{GC}=0$\,kpc plane) with a maximum distance of 25\,kpc from the node to the Galactic Center \citep{Marchetti2021}. In addition, a HiVel tracking back to the Galactic Center should intersect the Galactic midplane only once, and its backward-integral trajectory pass within 1\,kpc of the Galactic Center \citep{Liao2023}. For HiVels originating from the Sgr dSph, their backward-integral trajectories are expected to pass within the Sgr dSph's half-light radius of 2.587\,kpc \citep{McConnachie2012}. In the case of a HiVel originating from a globular cluster, it needs to pass within the Plummer scale radius of that cluster \citep{Vasiliev2021mn}. The probabilities were calculated by tallying the frequency of occurrence in the 4000 MC realizations conducted in this study.

\par
\begin{figure}[ht!]
\plotone{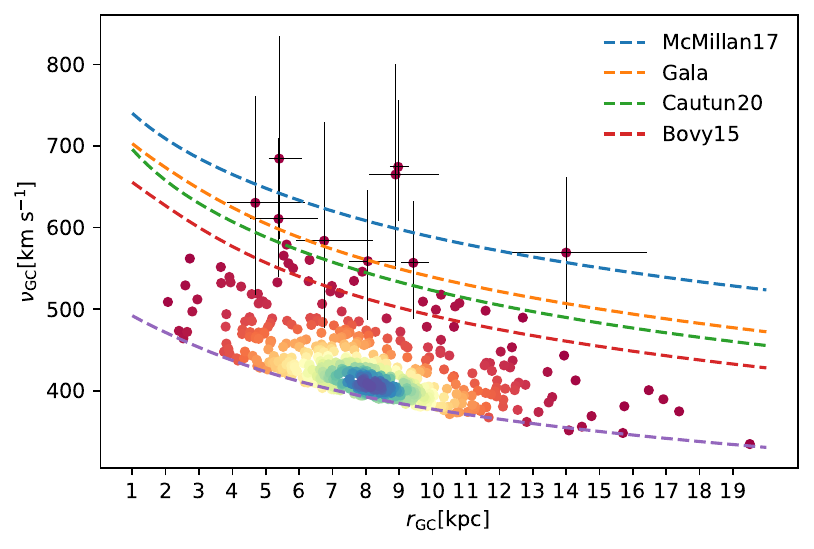}
\caption{The density distribution of total velocity $\nu\mathrm{_{GC}}$ vs. Galactocentric distance $r\mathrm{_{GC}}$ for the final sample 519 HiVels with $\nu\mathrm{_{GC}}$\,\textgreater\,$0.7\,\nu\mathrm{_{esc}}$, including 9 unbound candidates ($P_\mathrm{ub}$\,\textgreater\,0.5) with error bars under the {\tt\string Gala} \citep{gala} {\tt\string MilkyWayPotential}. The four different colored dashed lines are the escape velocity curves derived from the four different Milky Way potential models \citep{McMillan2017,gala,Cautun2020,bovy2015}, respectively. The purple dashed line represents 70\,\% of the escape velocity curve from the Gala. }\label{fig1}
\end{figure}

\begin{figure}[ht!]
\plotone{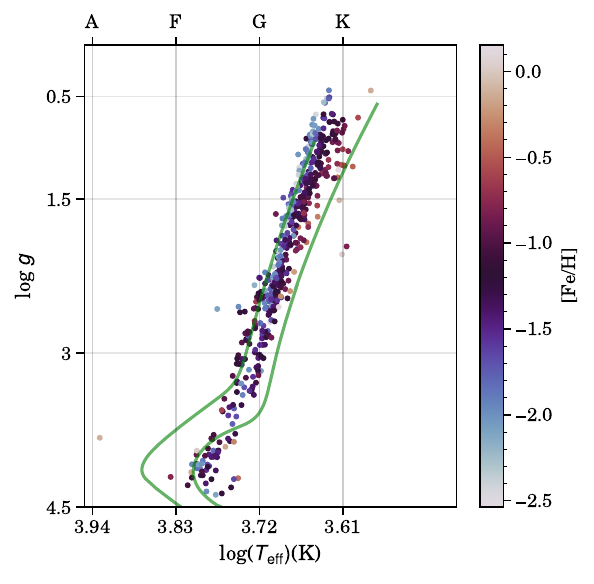}
\caption{A total of \textbf{513} HiVels from the final sample are shown in the lo\text{g}\,$\text{g}$ vs. $T_\mathrm{_{eff}}$ diagram, and the color is coded by metalicity. The right and left green lines represent MIST isochrones \citep{Dotter2016,Choi2016,Paxton2011,Paxton2013,Paxton2015} with [M/H]\,=\,$-0.5$, log(Age)\,[year]\,=\,9.8\,(100\,Myr), and [M/H]\,=\,$-2$, log(Age)\,[year]\,=\,9.8\,(100\,Myr), respectively.  } \label{fig2} 
\end{figure}

 The HiVels sample is defined as stars with total velocities ($\nu\mathrm{_{GC}}$) in the Galactocentric restframe exceeding 70\% of the escape velocity ($\nu\mathrm{_{GC}}$\,\textgreater\,$0.7\,\nu\mathrm{_{esc}}$). From this definition, a final sample of 519 HiVels with astrometric parameters and radial velocities was derived, including 9 unbound \textbf{candidates} ($P\mathrm{_{ub}}$\,$>$\,0.5). The density distribution of these stars in $\nu\mathrm{_{GC}}$ and $r\mathrm{_{GC}}$ is shown in Figure~\ref{fig1}. The number of HiVels is the highest at a distance of 8.1\,kpc and gradually decreases on both sides. This aligns with the observed number of stars. As the Sun is positioned at a distance of 8.12\,kpc \citep{GRAVITY2018A&A...615L..15G}, the count of observed stars gradually diminishes towards the outer regions. We also plot the Hertzsprung–Russell diagram of these stars' astrophysical parameters ($T\mathrm{_{eff}}$, lo\text{g}\,\textsl{g}, [Fe/H]) in Figure~\ref{fig2}. Most of them are the metal-poor late-type giants with [Fe/H]\,$<$\,$-$0.8\,dex and $T\mathrm{_{eff}}$\,$<$\,7300\,K.

\section{The origin of HiVels}
\label{4}

\subsection{Classification of HiVels}\label{2.1}

\par Table~\ref{tab1} gives the classification criteria of HiVels. Detailed information about 9 unbound stars (8 for the first time, and 2 reported
by \citet{Marchetti2021} and \citet{Li.Yinbi2021}) is provided in Table~\ref{tab4}. Two of these stars exhibit retrograde motion relative to the direction of the MW's rotation ($\nu\mathrm{_{\phi}}$\,$<$\,0), and their trajectories do not track back to the Galactic disk ($P\mathrm{_{disk}}$\,$<$\,0.5), suggesting a possible extragalactic origin. Furthermore, 57 bound HiVels ($P\mathrm{_{ub}}$\,$<$\,0.5) exhibit prograde motion and cluster at high metallicity, and their trajectories track back to the galactic disk ($P\mathrm{_{disk}}$\,$>$\,0.5), indicating they are likely the runaway stars (RSs) that form in the disk and then are ejected into the halo but not the typical OB type runaway stars. The remaining 453 HiVels are classified as high-velocity halo stars that may form within the MW or have extragalactic origins. 

\begin{deluxetable*}{ccccc}
\tablecaption{The classification criteria of HiVels \label{tab1}}
\tablehead{
Class & $P\mathrm{_{ub}}$  & $P\mathrm{_{disk}}$ & $\nu\mathrm{_{\phi}}$ & MBH ejection origin\\
 }
\startdata
Hypervelocity star (HVS)    & $ >0.5 $  & \dots  & \dots  & True\\
Hyper-runaway star (HRS)   & $ >0.5 $ & $ >0.5 $  & $>-276.7\times\mathrm{[Fe/H]}-97.78$ & False \\
Runaway star (RS)     & $<0.5$   & $>0.5$  & $>-276.7\times\mathrm{[Fe/H]}-97.78$ &  False  \\
High velocity halo star   & $<0.5$   & \dots  & \dots &  \dots \\
\enddata
\tablecomments{The formula $\nu\mathrm{_{\phi}}=-276.7\times\mathrm{[Fe/H]}-97.78$, proposed by \citet{LiQZ2023}, serves as an empirical demarcation line for distinguishing between the disk stars and the halo stars. $P\mathrm{_{ub}}$ and $P\mathrm{_{disk}}$ represent the probability of exceeding the Galactic escape velocity curves and the probability of tracking back to the Galactic disk, respectively. }
\end{deluxetable*}

\begin{deluxetable*}{ccccccccc}
\tablecaption{Basic parameters for the 9 unbound candidates}\label{tab4}
\tablehead{
Gaia DR3 source\_id & R.A.  & Decl.  & $rv$ & $d$ &  $\nu\mathrm{_{GC}}$ & $P\mathrm{_{ub}}$ & $P\mathrm{_{disk}}$ & $\nu\mathrm{_{\phi}}$\\
 & (deg)   &  (deg)  & (km\,s$^{-1}$) & (kpc) & (km\,s$^{-1}$)&  & & (km\,s$^{-1}$) }
\startdata
1309092223502856576 & 257.4199 & 29.9771 & $-487\pm1$ & $16.12_{-2.24}^{+3.00}$ & $570_{-70}^{+92}$ & 0.729 & 0.002 & $145_{-112}^{+142}$\\
1204061267883975040 & 238.1112 & 20.1969 & $-263\pm1$ & $10.36_{-1.36}^{+1.77}$ & $652_{-103}^{+149}$ & 0.825 & 0.007 & $-186_{-77}^{+185}$ \\
1297316350890352000 & 248.1776 & 21.2721 & $-335\pm1$ & $9.60_{-1.05}^{+1.33}$ & $559_{-72}^{+88}$ & 0.505 & 0.045 & $-26_{-80}^{+75}$\\
4263403134667526144 & 292.1825 & -0.0883 & $-25\pm2$ & $8.44_{-1.04}^{+1.48}$ & $694_{-108}^{+142}$ & 0.793 & 1 & $276_{-114}^{+176}$\\
6098873935647575552 & 220.6411 & -44.5675 & $163\pm1$ & $10.72_{-1.43}^{+1.98}$ & $582_{-106}^{+147}$ & 0.530 & 0.999 & $345_{-145}^{+200}$\\
6241679678390298880 & 236.9944 & -20.5118 & $-305\pm1$ & $10.90_{-1.31}^{+1.61}$ & $616_{-103}^{+171}$ & 0.549 & 0.043 & $574_{-144}^{+102}$ \\
6235773131986833792 & 239.3821 & -24.9635 & $-156\pm1$ & $10.78_{-1.55}^{+1.92}$ & $620_{-103}^{+140}$ & 0.549 & 0.972 & $592_{-210}^{+163}$\\
\hline
1383279090527227264 & 240.3373 & 41.1668 & $-185\pm1$ & $6.59_{-0.52}^{+0.61}$ & $676_{-68}^{+80}$ & 0.967 & 1 & $-487_{-37}^{+39}$ \\
1375165725506487424 & 231.8486 & 36.0344 & $-86\pm1$ & $7.79_{-0.70}^{+0.83}$ & $553_{-65}^{+79}$ & 0.546 & 1 & $-319_{-26}^{+21}$\\
\enddata
\tablecomments{The last two rows list the stars that have been reported by \citet{Marchetti2021,Li.Yinbi2021}.}
\end{deluxetable*}

\subsection{A HiVel track back to the Galactic Center}\label{2.2}

\par We found a HiVel that crosses over the Galactic midplane ($Z\mathrm{_{GC}}$\,$=$\,0\,kpc plane) only once and has a backwards-integrated trajectory passing within 1\,kpc of the Galactic Center. Table~\ref{tab2} shows the detailed information. Since the Galactic field is a classically collisionless system, particularly at the high speeds typical of HiVels, we neglect the perturbative influence of other bodies on these orbits. According to the Hills mechanism ejection probability proposed by \citet{Bromley2006}, the HiVel is not close enough to the Galactic Center, making it unlikely to be ejected by this mechanism. We computed the trajectories of 150 globular clusters from \citet{Vasiliev2019} using proper motions from \citet{Vasiliev2021mn}, but this star does not pass within their Plummer scale radius. Interestingly, the star is similar in metallicity and age to the VVV\,CL002 ([Fe/H]\,$=-0.4\pm0.2$, age\,$>$\,6.5\,Gyr) \citep{Bidin2011}, which is the closest globular cluster to the Galactic Center \citep{Minniti2021}. However, due to the lack of radial velocity information for this globular cluster, it is unknown whether the star passed within its Plummer scale radius.

\begin{deluxetable*}{ccc}
\tablecaption{Detailed information about the star with probable origin in the center of the Galaxy. \label{tab2}}
\tablehead{ Parameter & value  & unit  \\ }
\startdata
\textit{Gaia} DR3 source\_id    & 6236463526512162944   &   \\
GALAH DR3 sobject\_id   & 170603004101340  &    \\
$rv$    & $-144\pm0.3$   & km\,s$^{-1}$    \\
$d$   & $7.04_{-0.84}^{+1.05}$   & kpc   \\
$\nu_\mathrm{_{GC}}$ & $552_{-68}^{+89}$ & km\,s$^{-1}$ \\
$P\mathrm{_{cross}}$ & 0.761 & \\
$r\mathrm{_{closest}}$ & $0.79_{-0.34}^{+0.84}$ & kpc \\
\hline
$\mathrm{[Fe/H]}$ & $-0.58\pm0.09$ & dex\\
$T\mathrm{_{eff}}$ & $3891\pm93$  & K\\
lo\text{g}\,$\text{g}$  & $0.71\pm0.54$  & dex \\
Age  & 6.8767 &  Gyr \\
Mass & 1.07 & $M_\odot$  \\
\enddata
\tablecomments{$P\mathrm{_{cross}}$ and $r\mathrm{_{closest}}$ are the probability of crossing over the Galactic midplane only once and the distance of the backward-integral trajectory closest to the Galactic center, respectively. The last five rows of data come from GALAH DR3.}
\end{deluxetable*}

\subsection{HiVels originated from the Sgr dSph and the LMC }\label{2.3}

\begin{figure}[ht!]
\plotone{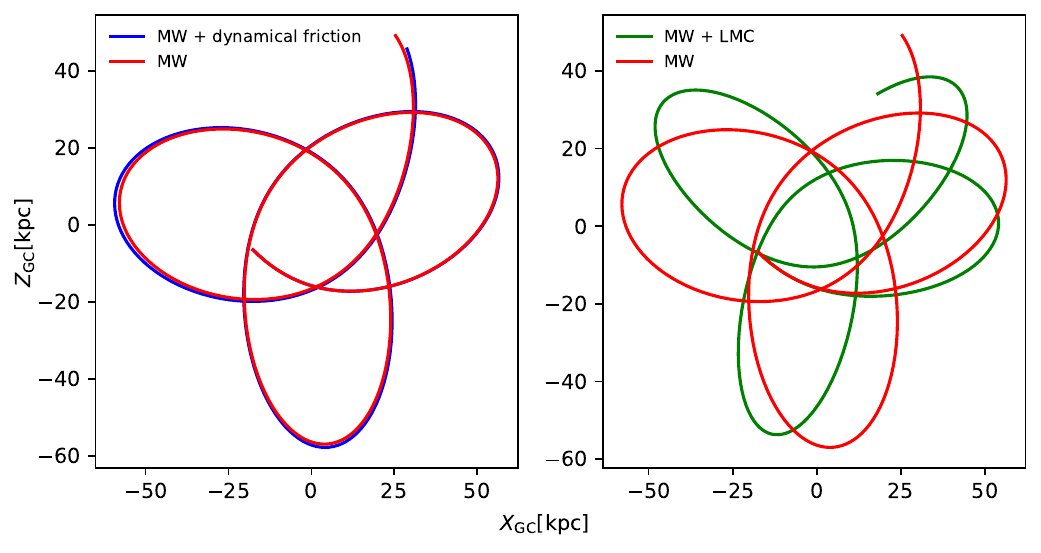}
\caption{The left panel illustrates the orbits without dynamical friction from the MW on the Sgr dSph and with dynamical friction, showing minimal differences. In contrast, the right panel shows the orbits without and with the potential of the LMC, displaying significant differences in the orbits caused by the LMC's gravitational potential.}\label{fig3} 
\end{figure}

In our sample of HiVels, three HiVels (2 for the first time, and 1 reported by \citet{Lihf2022}; \citet{LiQZ2023}) show potential origin from the Sgr dSph. Their backward-integrated trajectories can track back to the Sgr dSph ($P\mathrm{_{Sgr}}$\,$>$\,0.5) and other parameters are listed in Table~\ref{tab5}. The chemical abundance ([Fe/H]$=-1.12\pm0.02$, [$\alpha$/Fe]$=0.27\pm0.01$) of \textit{Gaia} DR3 3702750168409332864 is consistent with the members stars of the Sgr stream which are taken from an analysis based on the LAMOST K gaints \citep{Huang2021,Yang2019}. The color-magnitude (BP-RP\,$=$\,0.96, $M\mathrm{_G}$\,$=$\,0.21) of Gaia DR3 3653273996289384832 is most similar to both the Sgr and Sgr stream \citep[e.g.,][]{Lihf2022,Vasiliev2020,Vasiliev2021} 

\begin{figure}[ht!]
\plotone{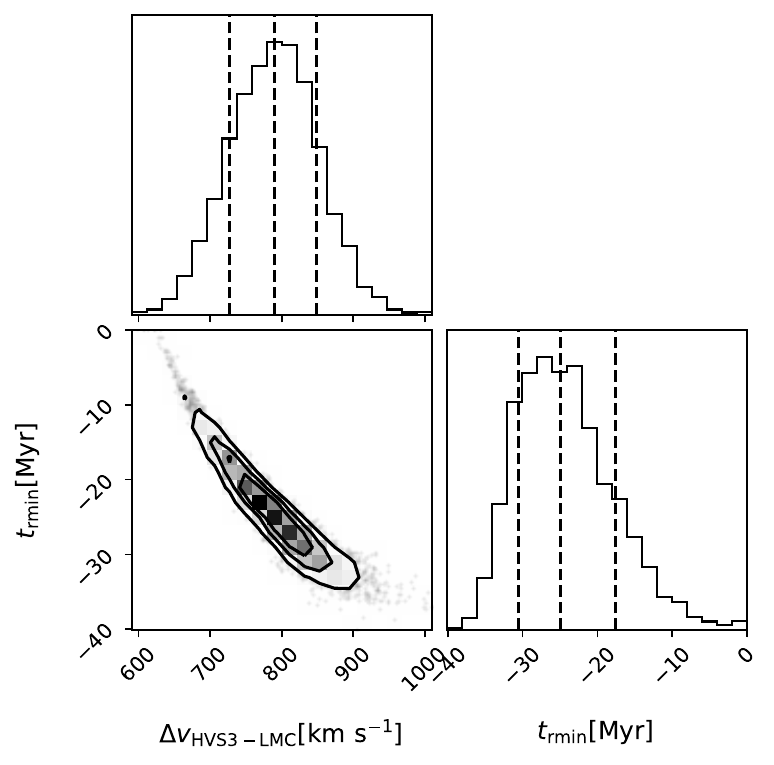}
\caption{Distribution of the velocity and time at closest approach to LMC. The dashed lines show the 16,50,84 percentiles of 1D distribution. The velocity is $790_{-63}^{+59}$\,km\,s$^{-1}$. The backward time is $24.9_{-7.3}^{+5.7}$\,Myr. }\label{fig5}
\end{figure}

It is evident that the LMC significantly influences the orbit of the Sgr dSph as shown in Figure~\ref{fig3}. As the HiVels that reported to originate from the Sgr dSph \citep{Lihf2022,Huang2021,LiQZ2023} did not consider the impact or the dynamical friction from the MW on the LMC, we recalculated the probabilities $P\mathrm{_{Sgr}}$ of their trajectories passing within the half-light radius of the Sgr dSph. Among them, the HiVels with $P\mathrm{_{Sgr}}$\,$>$\,0.5 are listed in Table~\ref{tab5}. We also calculated the trajectory of HVS3 \citep{Erkal2019} using the proper motion provided by \textit{Gaia} DR3, and related parameters are listed in Table~\ref{tab3}. HVS3 has a $37\%$ probability passing within 5\,kpc of the LMC, and we show contours of the time of closest approach to the LMC and the velocity during this closest approach in Figure~\ref{fig5}.

\begin{deluxetable*}{ccc}
\tablecaption{The related parameters of HVS3 \label{tab3}}
\tablehead{ Parameter & value  & unit  \\ }
\startdata
R.A.    & $04:38:12.8$   &   \\
Decl.  & $-54:33:12$  &    \\
$rv$    & $723\pm3$   & km\,s$^{-1}$    \\
$d$   & $61\pm10$   & kpc   \\
\hline
$\mu _{\alpha*}$ & $0.853\pm0.049$ & mas\,yr$^{-1}$ \\
$\mu _{\delta }$ & $1.614\pm0.061$  & mas\,yr$^{-1}$ \\
Corr($\mu _{\alpha*}$,\,$\mu _{\delta }$)  & -0.086  &  \\
\botrule
\enddata
\tablecomments{The first four rows refer to \citet{Erkal2019}, and the last three rows are from \textit{Gaia} DR3. }
\end{deluxetable*}

\begin{longrotatetable}
\begin{deluxetable*}{llllllllll}
\tablecaption{HiVels originating from the Sgr dsph}\label{tab5}
\tablehead{
Gaia DR3 source\_id  &   rv  & $d$ &  $\nu\mathrm{_{GC}}$ & $P\mathrm{_{Sgr}}$ & $G-$magnitude$^{a}$ & BP-RP$^{a}$ & [Fe/H] &  [$\alpha$/Fe] 
 & Reference\\
\dots   & (km\,s$^{-1}$)  & (kpc) &  (km\,s$^{-1}$) &  \dots & mag & mag & dex & dex & \dots }
\startdata
3702750168409332864 & $465\pm3$ & $2.47_{-0.13}^{+0.15}$ & $410_{-2}^{+3}$ & 0.64 & 14.4 & 0.96 & $-1.23\pm0.02$$^{b}$ & $0.27\pm0.01$$^{b}$ & (1) \\
3653273996289384832 & $337\pm4$ & $5.39_{-0.51}^{+0.62}$ & $505_{-28}^{+39}$ & 0.96 & 13.9 & 0.96 & $-1.02\pm0.09$$^{b}$ & \dots & (1) \\
3793871060689209984 & $507\pm1$ & $2.30_{-0.10}^{+0.12}$ & $419_{-1}^{+1}$ & 1.0 & 11.3 & 1.18 & $-1.02\pm0.28$$^{c}$ & \dots & (1)(2)(3) \\
564961308084113152 & $-70\pm3$ & $2.40_{-0.06}^{+0.06}$ & $417_{-12}^{+12}$ & 1.0 & 12.3 & 1.18 & \dots & \dots & (2) \\
1805174788871484800 & $-368\pm3$ & $3.80_{-0.20}^{+0.23}$ & $415_{-14}^{+17}$ & 0.92 & 12.6 & 0.89 & \dots & \dots & (2) \\
4423441557511145088 & $226\pm2$ & $8.02_{-0.96}^{+1.18}$ & $422_{-38}^{+52}$ & 0.60 & 12.8 & 1.54 & \dots & \dots & (2) \\
6716076404123420288 & $-278\pm4$ & $3.06_{-0.13}^{+0.14}$ & $488_{-15}^{+15}$ & 0.93 & 12.3 & 1.29 & \dots & \dots & (2) \\
6734684023516484096 & $-334\pm0$ & $8.50_{-0.88}^{+1.11}$ & $501_{-33}^{+45}$ & 0.58 & 13.1 & 1.50 & $-1.21\pm0.01$ & $0.22\pm0.01$ & (2) \\
383206057417413248 & $-313\pm13$ & $1.48\pm0.07$ & $443\pm19$ & 1.0 & 15.3 & 0.76 & \dots & \dots & (3)\\
1282270187101294592 & $184\pm13$ & $2.53\pm0.34$ & $439\pm55$ & 0.57 & 16.69 & 0.72 & \dots & \dots & (3)\\
1962742265496273920 & $-318\pm3$ & $1.72\pm0.27$ & $452\pm66$ & 0.62 & 17.55 & 1.31 & \dots & \dots & (3)\\
\enddata
\end{deluxetable*}
\tablecomments{Reference: (1)\,this work; (2)\citet{Lihf2022}; (3)\citet{LiQZ2023}. [$^a$]: $G-$magnitude and BP-RP from \textit{Gaia} DR3.  [$^b$]: Data from LAMOST DR10.  [$^c$]: Data from RAVE DR6. }
\end{longrotatetable}

\section{Summary and Discussion}
\label{5}

\par In this study, we identify a total of 519 high-velocity stars (HiVels) through a cross-matching process between the \textit{Gaia} DR3 and other Large-scale Galactic Surveys.  A catalog of the properties of these stars is available on China-VO: \dataset[doi:10.12149/101304]{https://nadc.china-vo.org/res/r101304/}. These HiVels consist of 9 unbound candidates under the {\tt\string Gala} {\tt\string MilkyWayPotential}, 57 runaway stars (not the typical OB type runaway stars), and 453 high-velocity halo stars. In our sample, 78 HiVels have been reported by \citet{Li.Yinbi2021}, a candidate HiVel originating from the Sgr dSph has been reported by \citet{Lihf2022}, and 440 have been reported for the first time.

\par The \textit{Gaia} \textit{G}-band magnitudes of these HiVels are less than 15 and majority of these stars are metal-poor late-type giants. Although we employ geometric distances rather than photogeometric or photo-astrometric distances, they are very similar at \textit{G}\,$<$\,15 as shown in \citet{Anders2022}. The backwards-integraed trajectory of a HiVel can track back to the Galactic Center but there is insufficient evidence for the VVV\,CL002 origin. Notably, based on the color-magnitude or chemical analysis, as well as the trajectories, three HiVels could be the member stars of the Sgr dSph. If the LMC has not been taken into account in the orbital analysis, we still obtain the same results for the four stars, as they are far from the LMC (over 50\,kpc) and have a short time to their pericenter (57\,Myr ago). Considering the impact of the LMC, only a small number of reported HiVels pass within the half-light radius of the Sgr dSph. This is because the LMC has a significant impact on the orbit of the Sgr dSph and HiVels originating from the Sgr dSph are required to pass within the half-light radius of the Sgr dSph. Using the \textit{Gaia} DR3's proper motion, HVS3 may still come from the LMC. In the near future, we anticipate the radial velocity of VVV\,CL002 to be observed, which will enable us to calculate its orbit to determine the origin of the HiVel tracking back to the Galactic Center. We also expect that the future surveys will provide a rich source of more HiVels to provide new insights for understanding the presence of an extreme dynamic in the Galaxy.

\section*{Acknowledgments}

\par   We thank especially the referee for insightful comments and suggestions, which have improved the paper significantly. We are grateful to Jo Bovy for his helpful advice on orbital integration. This work was supported by the National Natural Science Foundation of China (NSFC Nos: 12090040, 12090044,
11973042, 11973052, and 11873053).  
It was also supported by the Fundamental Research Funds for the Central Universities and the National Key R\&D Program of China No. 2019YFA0405501.  
This work has made use of data from the European Space Agency (ESA) mission
{\it Gaia} (\url{https://www.cosmos.esa.int/gaia}), processed by the {\it Gaia}
Data Processing and Analysis Consortium (DPAC,
\url{https://www.cosmos.esa.int/web/gaia/dpac/consortium}). Funding for the DPAC
has been provided by national institutions, in particular the institutions
participating in the {\it Gaia} Multilateral Agreement.

\par This work has made use of data from the European Space Agency (ESA) mission {\it Gaia} (\url{https://www.cosmos.esa.int/gaia}), processed by the {\it Gaia} Data Processing and Analysis Consortium (DPAC, \url{https://www.cosmos.esa.int/web/gaia/dpac/consortium}). Funding for the DPAC has been provided by national institutions, in particular the institutions participating in the {\it Gaia} Multilateral Agreement.

\par This work made use of the Third Data Release of the GALAH Survey. This paper includes data that has been provided by AAO Data Central (\url{datacentral.aao.gov.au}). It also made use of the Sixth Data Release of the RAVE Survey. Guoshoujing Telescope (the Large Sky Area Multi-Object Fiber Spectroscopic Telescope LAMOST) is a National Major Scientific Project built by the Chinese Academy of Sciences. Funding for the project has been provided by the National Development and Reform Commission. Funding for the Sloan Digital Sky Survey IV has been provided by the Alfred P. Sloan Foundation, the U.S. Department of Energy Office of  Science, and the Participating  Institutions. SDSS-IV acknowledges support and resources from the Center for High Performance Computing at the  University of Utah. The SDSS website is www.sdss.org. Based on data acquired at the Anglo-Australian Telescope. We acknowledge the traditional owners of the land on which the AAT stands, the Gamilaraay people, and pay our respects to elders past and present.

\appendix
\section{Catalog Format}

Table \ref{tab:catalog} describes the format of the full table for the 519 HiVels with a total velocity in the Galactocentric restframe greater than 70\% of the local escape velocity.
It is available in a machine-readable format in the online Journal and in China-VO.

\startlongtable

\begin{deluxetable*}{ccl}
\tablecaption{Catalog Format \label{tab:catalog}}
\tablehead{
\colhead{Label} &
\colhead{Unit} &
\colhead{Description}
}
\startdata
designation    & ---    &  Unique source designation, Gaia DR3  \\
RAdeg           & deg    &  Right ascension, decimal degrees, Gaia DR3 (ICRS at Epoch=2016.0)         \\
DEdeg          & deg    &  Declination, decimal degrees, Gaia DR3 (ICRS at Epoch=2016.0)                               \\
plx            & mas    &  Parallax, Gaia DR3          \\
e\_plx         & mas    &  Error of plx, Gaia DR3      \\
plx-zp         & mas    &  Parallax zero-point correction, after \citet{Lindegren2021a}         \\
pmRA           & mas.yr-1  &  Proper motion in right ascension direction, Gaia DR3                                \\
e\_pmRA        & mas.yr-1  &  Error of pmRA             \\
pmDE           & mas.yr-1  &  Proper motion in declination direction, Gaia DR3                                            \\
e\_pmDE        & mas.yr-1  &  Error of pmDE           \\
plx\_pmra-corr  & ---    &  Correlation between parallax and proper motion in right ascension, Gaia DR3                \\
plx\_pmdec-corr & ---    &  Correlation between parallax and proper motion in declination, Gaia DR3                    \\
pmra\_pmdec-corr & ---   &  Correlation between proper motion in right ascension and in declination, Gaia DR3          \\
ruwe             & ---   &  Renormalised unit weight error, Gaia DR3                                                    \\
Gmag             & mag   &  G-band mean magnitude, Gaia DR3 [phot\_g\_mean\_mag]                                        \\
bp-rp            & mag   &  BP-RP color, Gaia DR3 [phot\_bp\_mean\_mag - phot\_rp\_mean\_mag]             \\
cross-match-sur  & ---   &  Cross-match the name of survey id\\
surveyid         & ---   &  Unique ID in the survey [survey\_id]\\
Teff             & K     &  Effective temperature, the cross-match survey                                           \\
e\_Teff          & K     &  Error of Teff              \\
logg             & [cm.s-2]   &  Surface gravity, the cross-match survey                     \\
e\_logg          & [cm.s-2]   &  Error of logg            \\
FeH              & [-]    &  Metallicity, the cross-match survey \\
e\_FeH           & [-]    &  Error of metallicity        \\
a/FeH            & [-]    &  Alpha abundance, the cross-match survey        \\
e\_a/FeH         & [-]       &  Error of a/FeH        \\
RVel             & km.s-1    &  Weighted mean radial velocity, Equation 1    \\
e\_RVel          & km.s-1    &  Error of radial velocity \\
GGDrlen          & ---       &  Corresponding to L in equation 2, after \citet{Bailer-Jones2021}       \\
GGDalpha         & ---       &  Corresponding to $\alpha$ in equation 2, after \citet{Bailer-Jones2021}         \\
GGDbeta          & ---       &  Corresponding to $\beta$ in equation 2, after \citet{Bailer-Jones2021}       \\
Dist             & kpc       &  Heliocentric distance (after \citet{Du2019}) \\
e\_Dist          & kpc       &  Lower uncertainty on Dist   \\
E\_Dist             & kpc       &  Upper uncertainty on Dist    \\
RGC              & kpc       &  Galactocentric distance (after \citet{Du2019})                                                \\
e\_RGC         & kpc       &  Lower uncertainty on RGC  \\
E\_RGC         & kpc       &  Upper uncertainty on RGC    \\
VGC            & km.s-1    &  Galactocentric total velocity (after \citet{Du2019})      \\
e\_VGC         & km.s-1    &  Lower uncertainty on VGC\\
E\_VGC        & km.s-1    &  Upper uncertainty on VGC\\
Vphi           & km.s-1    &  Rotation velocity, this work\\
P-ub             & ---       &  Unbound possibility, this work \\
P-disk           & ---       &  Probablity of the star tracking back to the disk , this work         \\
P-sgr            & ---       &  Probablity of the star tracking back to the Sgr, this work     \\
var-flag          & ---       &  Flag indicating if variability was identified in the photometric data,  \\
 &  &        Gaia DR3 [phot\_variable\_flag] \\
Pstar             & ---       &  Probability from DSC-Combmod of being a single star,  \\
    &         & Gaia DR3, [classprob\_dsc\_combmod\_star] \\
RVel-DR3         & km.s-1    &  Gaia DR3 radial velocity  \\
e\_RVel-DR3        & km.s-1   &  Error of RVel-DR3  \\
\enddata
\tablecomments{Table \ref{tab:catalog} is published in its entirety in the machine-readable format. 
A portion is shown here for guidance regarding its form and content. 
It is also available on China-VO: \dataset[doi:10.12149/101304]{https://nadc.china-vo.org/res/r101304/}}
\end{deluxetable*}

\clearpage\newpage

\bibliography{ljw}
\bibliographystyle{aasjournal}

\end{document}